\documentclass[a4paper,onecolumn,11pt,accepted=2026-04-27]{quantumarticle}
\pdfoutput=1
\usepackage[utf8]{inputenc}
\usepackage[english]{babel}
\usepackage[T1]{fontenc}
\usepackage{amsmath}
\usepackage{hyperref}
\usepackage{booktabs}
\usepackage{tikz}
\usepackage{lipsum}
\usepackage{quantikz}
\usepackage[most]{tcolorbox}
\usepackage[numbers,sort&compress]{natbib}

\begin{document}

\title{Efficient implementation of single particle Hamiltonians in exponentially reduced qubit space}

\author{Martin Plesch}
\affiliation{Institute for Physics, Slovak Academy of Sciences, Bratislava, Slovakia}
\affiliation{Matej Bel University, Banská Bystrica, Slovakia}
\author{Martin Friák}
\affiliation{
 Institute of Physics of Materials, v.v.i., Czech 
Academy of Sciences, Žižkova 22, CZ-616~00, Brno, Czech Republic
}
\affiliation{Department of Condensed Matter Physics, Faculty of Science, Masaryk University, Kotlářská 2, CZ-611~37, Brno, Czech Republic}
\author{Ijaz Ahamed Mohammad}%
\email{fyziijaz@savba.sk}
\affiliation{Institute for Physics, Slovak Academy of Sciences, Bratislava, Slovakia}
\maketitle

\begin{abstract}
Current and near-term quantum hardware is constrained by limited qubit counts, circuit depth,
and the high cost of repeated measurements. We address these challenges for solid-state
Hamiltonians by introducing a logarithmic-qubit encoding that maps a system with $N$ physical
sites onto only $\lceil \log_2 N \rceil$ qubits while maintaining a clear correspondence with the
underlying physical model. Within this reduced register, we construct a compatible variational
circuit and a Gray-code-inspired measurement strategy whose number of global settings grows
only logarithmically with system size. To quantify the overall hardware load, we introduce a
volumetric efficiency metric that combines the number of qubits, circuit depth, and the number
of measurement settings into a single measure, expressing the overall computation costs. Using
this metric, we show that the total space--time sampling volume required in a variational loop
can be reduced dramatically from $N^2$ to $(\log N)^3$ for a hardware-efficient ansatz, allowing
an exponential reduction in time and size of the quantum hardware. These results demonstrate
that large, structured solid-state Hamiltonians can be simulated on substantially smaller quantum
registers with controlled sampling overhead and manageable circuit complexity, extending the
reach of variational quantum algorithms on near-term devices.
\end{abstract}

\section{\label{sec:intro}Introduction}
Determining the energy spectrum of molecules and condensed-matter systems lies at the heart of quantum chemistry and many-body physics. 
The problem can be formally expressed as finding the eigenvalues and eigenstates of a high-dimensional Hamiltonian, whose complexity grows exponentially with the number of interacting particles and orbitals. 
As a result, obtaining accurate electronic energies and wavefunctions remains one of the major computational bottlenecks in modern quantum science \cite{helgaker2013molecular,szabo2012modern}.

Classical computational approaches, such as Full Configuration Interaction (FCI) \cite{knowles1984new}, Coupled-Cluster (CC) theory \cite{bartlett2007coupled}, and Density Functional Theory (DFT) \cite{parr1995density}, have achieved remarkable success in describing a wide range of chemical and physical systems. 
However, their computational cost grows rapidly with system size. 
Even the widely used CCSD (Coupled-Cluster Singles and Doubles) method, which captures most of the electron correlation by truncating excitations to the single- and double-excitation manifold, scales as $\mathcal{O}(N^6)$ in the number of orbitals \cite{bartlett2007coupled}. 
For larger or strongly correlated systems, such methods become intractable, necessitating approximate or model-specific simplifications.

Classically, eigenvalues and eigenstates are typically computed with sparse tools from linear algebra (e.g. the Lanczos algorithm, also known as the recursion method~\cite{teng2011efficient}). This is particularly relevant for semiconductors and nanostructures such as quantum dots which are often well described in a single-particle model~\cite{8d5x-779h, lu2006moments, mittelstadt2022modeling}. In addition to scaling challenges, classical iterative eigensolvers such as the Lanczos algorithm suffer from numerical instability. 
In finite precision, the orthogonality of the Krylov subspace vectors gradually degrades, resulting in spurious or unstable eigenvalues~\cite{paige1976error}. 
Reorthogonalization techniques--such as Gram–Schmidt or block-based methods--can restore numerical stability but at the expense of significant memory overhead and increased runtime, especially at high spectral resolution~\cite{parlett1979lanczos, ozaki2006n}. 
This trade-off between precision and scalability fundamentally limits classical simulations of large Hamiltonians.

These limitations motivate the exploration of quantum algorithms as a novel and promising approach for computing molecular and material energy spectra. 
Quantum hardware provides a natural platform to represent and manipulate exponentially large Hilbert spaces with polynomial resources. 
In particular, Variational Quantum Algorithms (VQAs)~\cite{cerezo2021variational} have emerged as a promising class of methods that combine parameterized quantum circuits with classical optimization. 
Among them, the Variational Quantum Eigensolver (VQE)~\cite{peruzzo2014variational, fedorov2022vqe, kandala2017hardware, tilly2022variational,mohammad2024meta,mihalikova2022cost} stands out as a practical candidate for near-term devices: it minimizes the expectation value of a Hamiltonian to approximate ground state using shallow ansätze. 
For higher-energy states, extensions such as Variational Quantum Deflation (VQD)~\cite{higgott2019variational} further refine the energy spectrum by introducing orthogonality penalties against previously found eigenstates.

Similar to classical truncation schemes such as CCSD, one can restrict the quantum Hilbert space to a physically motivated subset of configurations. 
In this work, we focus on the single-particle Hamiltonian, which describes a single particle delocalized across $N$ orbitals or lattice sites. 
This model naturally maps onto a tight-binding (TB) Hamiltonian of dimension $N$, which models electrons in a crystal using a localized orbital basis, with diagonal terms corresponding to on-site energies and off-diagonal terms representing hopping amplitudes between sites. TB is used to model band structures with a small set of physically informed parameters. 
Parameters can be obtained empirically by fitting band edges and effective masses (empirical TB)~\cite{jancu1998empirical, phan2024empirical, vogl1983semi} or derived from first-principles frameworks such as density functional theory~\cite{tan2013empirical}.
Despite its apparent simplicity, this approximation retains essential physical structure and provides a meaningful testbed for studying variational quantum algorithms under realistic hardware constraints. 

While single-particle and tight-binding Hamiltonians do not by themselves constitute classically hard problems, they offer a useful benchmark for developing efficient quantum protocols. We view the present work as meaningful progress toward that broader objective, developing tools and procedures to be used on more complex systems, when quantum devices will become available in such a scale.

This single-particle approach for VQD was formulated in \cite{sherbert2021systematic}, and subsequent work by the present authors \cite{krejvci2025minimum} significantly reduced the measurement overhead from an explicit \(O(N)\) scaling to a constant number of \textit{three} global measurement settings, independent of the system size. In the present work, we take this approach a step further by showing how to realize the same class of Hamiltonians on a logarithmic number of qubits, thereby reducing the required quantum resources even further. Although Hamiltonians of dimension \(N\) can, in principle, be represented using only \(\log_2 N\) qubits, practical and resource-efficient implementations have received relatively little attention in the literature~\cite{mcardle2020quantum,sherbert2021systematic,sherbert2022quantum}.Here, we close this gap in knowledge by providing a logarithmic-register variational approach that extends the reach of single-particle simulations to substantially larger system sizes on limited hardware.

To compare the efficiency of hybrid quantum–classical algorithms, we introduce a volumetric measure that captures the actual resource demand of such procedures. In classical computation, two basic parameters are considered: memory demand (number of bits) and time (number of operations). These parameters are mostly independent, and each independently limits the feasibility of executing the algorithm: the algorithm must both fit into the computer’s memory and complete within a bearable time.

In the quantum world, the situation is different. While the size of the device (number of qubits) is likewise a limiting parameter, the notion of time is more subtle. First, the output of a quantum computer is quantum, and converting it into a classical, human-readable form requires measurements. The number of incompatible measurement bases affects the efficiency of the algorithm, since more bases require more independent runs. Moreover, as most quantum devices can, unlike classical ones, perform many gates in parallel, the depth of the circuit (number of layers) is more relevant than the total number of operations. Finally, because the outcomes of quantum measurements are stochastic, many repetitions are needed; on large devices, these repetitions can be parallelized, speeding up the process. Based on these differences, we define a measure given by the product of the device size, circuit depth, and number of incompatible measurement settings, and use it to highlight the performance of our approach.

The paper is organized as follows.
In Section II, we outline the theoretical framework of the Variational Quantum Eigensolver (VQE) and introduce the specific Single Exitation Subspace (SES) Hamiltonians.
Section III discusses optimal state preparation strategies  on a logarithmic number of qubits for such Hamiltonians, while Section IV deals with measurement settings, highlighting how its structure enables a significant reduction in its number.
In Section V we show the exponential speedup of the presented algorithm in comparison with the previously used approchaes and the Section VI concludes our findings.

\section{Background}
\subsection{Variational Quantum Eigensolver}

The Variational Quantum Eigensolver (VQE)~\cite{peruzzo2014variational} is a hybrid quantum-classical algorithm designed to estimate the ground-state energy of a given Hamiltonian. Its relatively low circuit depth and reliance on classical optimization make it particularly suitable for near-term quantum devices.

VQE functions by preparing a parameterized quantum state $\ket{\psi(\boldsymbol{\theta})}$ using a variational ansatz, typically constructed from layers of parameterized single-qubit rotations and entangling gates. The expectation value of the Hamiltonian $H$ with respect to this state, $\langle \psi(\boldsymbol{\theta}) | H | \psi(\boldsymbol{\theta}) \rangle$, is estimated through quantum measurements and serves as the cost function for optimization.

A classical optimizer is employed to minimize this cost by iteratively updating the parameters $\boldsymbol{\theta}$, with the aim of converging to the ground state of the system. This hybrid loop continues until a convergence criterion is met, ideally resulting in a quantum state that closely approximates the true ground state of the target system.

In recent years, VQE has become a standard benchmark for variational quantum algorithms across various domains, including quantum chemistry, condensed matter physics, and combinatorial optimization~\cite{cerezo2021variational}. Nonetheless, its performance is strongly influenced by the choice of ansatz, the effectiveness of the classical optimization strategy \cite{mohammad2025hopso}, and the presence of quantum noise. A very important aspect is also the implementation of the physical Hamiltonian, addressed in the paper.

\subsection{Solid states SES Hamiltonians}
The Single Excitation Subspace (SES), equivalent to single-particle approximation in tight-binding method, refers to a constrained region of the full Hilbert space in which only one excitation is present across all qubits or quantum modes at any given time. Mathematically, the SES consists of all quantum states that are linear combinations of basis states with a single excitation, such as $\ket{100\ldots0}$, $\ket{010\ldots0}$, ..., $\ket{000\ldots1}$. This subspace is particularly relevant for simulating physical systems where the dynamics are dominated by a single particle or quasi-particle, such as an exciton, electron, or spin flip.

The SES approximation is powerful because it drastically reduces the dimensionality of the problem--from an exponentially large space of $2^n$ states for $n$ qubits to only $n$ states--making both classical and quantum simulations more tractable. This makes SES particularly useful for near-term quantum hardware, which is limited in qubit number and coherence time.

Although the SES describes a linear system, it remains relevant for electronic structure problems. In particular, the single-excitation subspace captures the physics of one-electron models, such as the calculation of molecular orbitals in the linear combination of atomic orbitals (LCAO) framework. The SES forms the basis for computing one-electron energy levels and can serve as a starting point for more advanced approaches. Even though the SES does not include electron–electron interactions, it provides a computationally efficient approximation to simulating the electronic properties of materials.

\subsubsection{SES Ansatz}

In an $N$-qubit system restricted to the Single Excitation Subspace (SES), 
the variational state must encompass all basis vectors $\{|e_j\rangle\}_{j=1}^N$ 
of Hamming weight one. A compact strategy to generate such states begins 
with a localized excitation, e.g.\ $|e_1\rangle = X_1|0\rangle^{\otimes N}$, 
and subsequently propagates it through the register by means of 
two-qubit entangling gates.  

We use family of gates, $\hat{A}_{j,j+1}(\beta_j,\gamma_j)$, 
that act on neighboring qubits and mix the amplitudes of $|e_j\rangle$ 
and $|e_{j+1}\rangle$ \cite{sherbert2021systematic}. By applying these gates sequentially across the chain, 
one constructs the trial state
\begin{equation}
|\psi(\boldsymbol{\theta})\rangle =
\left( \prod_{j=1}^{N-1} \hat{A}_{j,j+1}(\beta_j,\gamma_j) \right) X_0 |0\rangle^{\otimes N}.
\label{eq:ses_ansatz}
\end{equation}

After all $N-1$ gates have been applied, the register occupies a general 
superposition of single-excitation states,
\begin{equation}
|\psi\rangle = \sum_{j=1}^N \alpha_j |e_j\rangle,
\label{eq:ses_superposition}
\end{equation}
with coefficients $\alpha_j = |\alpha_j| e^{i\theta_j}$ capturing both 
amplitude ($|\alpha|$) and phase ($\theta$) of the excitation at site $j$.

\begin{figure*}[t]
    \centering
        \centering
        \begin{quantikz}[color=black,background color=green!25]
            \lstick{$q_{\rm 0}$} \ket{1} \quad & \gate[2]{\hat A_{\rm 0, 1}} & \ghost{H} & \ghost{H} & \ghost{H} & \qw \\
            \lstick{$q_{\rm 1}$} \ket{0} \quad & \ghost{H} & \gate[2]{\hat A_{\rm 1, 2} } & \ghost{H} & \ghost{H} & \qw \\
            \lstick{$q_{\rm 2}$} \ket{0} \quad & \ghost{H} & \ghost{H} & \ghost{H} & \ghost{H} & \qw \\
            \lstick{$\vdots$} \quad & \ghost{H} & \ghost{H} & \ \ldots\ & \ghost{H} & \qw \\
            \lstick{$q_{\rm {n - 1}}$} \ket{0} \quad & \ghost{H} & \ghost{H} & \ghost{H} & \gate[2]{\hat A_{\rm n - 1, n} } & \qw \\
            \lstick{$q_{\rm n}$} \ket{0} \quad & \ghost{H} & \ghost{H} & \ghost{H} & \ghost{H} & \qw \\
        \end{quantikz}
        \captionsetup{justification=justified} 
        \captionof{figure}{Circuit representation of the SES ansatz. The localized excitation is propagated across the register through a sequence of two-qubit entangling gates $\hat{A}_{j,j+1}$ acting on neighboring qubits.}
        \label{fig:ansatz}
\end{figure*}
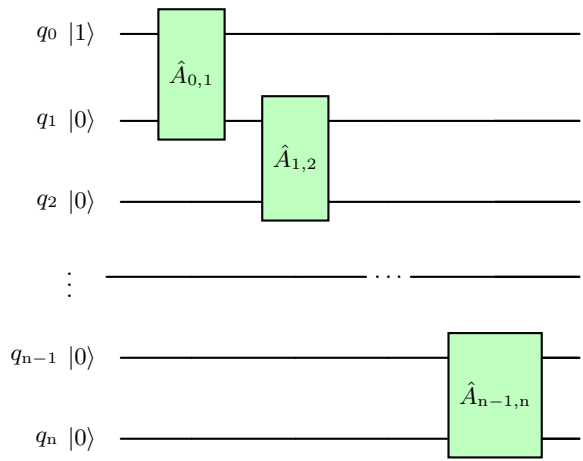

\begin{figure*}[t]

    \centering
    \resizebox{\textwidth}{!}{%
    \begin{quantikz}[color=black,background color=green!25]
    &\gate[2]{\hat A_{j,j+1}}& \\
    & \ghost{H} &
    \end{quantikz}=\begin{quantikz}
    & 
    \ctrl{1}\gategroup[2,steps=7,style={dashed,rounded
    corners,fill=green!25, inner
    xsep=2pt},background,label style={label
    position=below,anchor=north,yshift=-0.2cm}]{}  & \gate{R_z^{\dagger}(\gamma_{\rm j} + \pi)} & \gate{R_y^{\dagger}(\beta_{\rm j} + \pi/2)} & \targ{} & \gate{R_y(\beta_{\rm j} + \pi/2)}& \gate{R_z(\gamma_{\rm j} + \pi)} & \ctrl{1} &    \\
    &  \targ{} &&& \ctrl{-1} &&& \targ{} &
    \end{quantikz}
    }
        \vspace{0pt}
        \captionsetup{justification=justified} 
        \captionof{figure}{Implementation of the $\hat{A}_{j,j+1}$ gate used in the SES ansatz. Each block is expressed in terms of three CNOTs and a sequence of parameterized single-qubit rotations $R_y$ and $R_z$.}
    \label{fig:Agate}

\end{figure*}
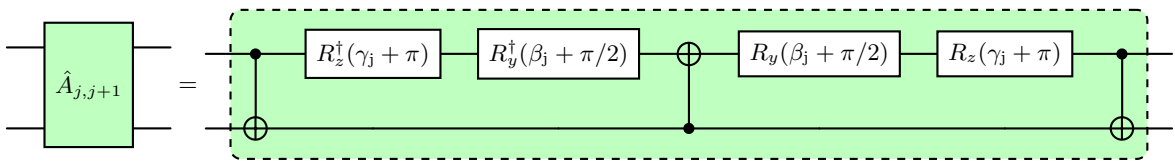

The Fig.~\ref{fig:ansatz} illustrates the sequential layout of the entangling gates, 
which effectively ``sweep'' the excitation across the system. Each gate $\hat{A}_{j,j+1}$ admits a standard decomposition into elementary operations: three two-qubit controlled gates (e.g.\ CNOTs), two $R_z$ rotations, and two $R_y$ rotations, as shown in Fig.~\ref{fig:Agate}.  

\subsubsection{Hamiltonian}
In the single-excitation subspace (SES), the system evolves within the sector of the Hilbert space that contains exactly one excitation. Each computational basis state in this subspace corresponds to a single qubit being in the excited state $\ket{1}$, with all other qubits remaining in the ground state $\ket{0}$. The general SES Hamiltonian, which is mathematically equivalent to the tight-binding Hamiltonian commonly used in solid-state physics, includes both on-site energy terms $h_{jj}$ and coherent hopping terms $h_{jk}$ between different sites. It can be written as:

\begin{equation}
H = \sum_{k=1}^{N} h_{kk} \ket{k}\bra{k} + \sum_{j \ne k} h_{jk} \ket{j}\bra{k},
\end{equation}
where $\ket{j}$ denotes the state with the excitation localized at site $j$.

To express this Hamiltonian in terms of Pauli operators acting on the full qubit space, we map the diagonal terms using the projector $\ket{1}_j\bra{1} = \frac{1 - Z_j}{2}$ and the off-diagonal hopping terms using the ladder operators: $\ket{j}\bra{k} = \sigma_j^+ \sigma_k^-$, where $\sigma_j^{\pm} = \frac{1}{2}(X_j \mp i Y_j)$. Decomposing these into Pauli matrices leads to the full qubit Hamiltonian \cite{sherbert2021systematic}:
\begin{align} \label{H}
H &= \sum_{k=1}^N \frac{h_{kk}}{2}(1 - Z_k) \notag \\[6pt]
&\quad + \sum_{j < k} \Bigg(
     \frac{\mathrm{Re}(h_{jk})}{2}\,(X_j X_k + Y_j Y_k) \notag \\[6pt]
&\qquad\qquad\;
   + \frac{\mathrm{Im}(h_{jk})}{2}\,(Y_j X_k - X_j Y_k)
   \Bigg).
\end{align}

This form is Hermitian and preserves the excitation number, ensuring that time evolution remains within the SES. It also provides a convenient basis for implementing SES dynamics on quantum hardware using variational approaches.

\section{State preparation}

While a constant number of measurement settings as previously shown in our \cite{krejvci2025minimum} constitutes a very compelling reduction on the number of runs of the quantum device needed, the algorithm still relies on a large number of qubits $N$ and a long series of $O(N)$ subsequent gates. Here we present a different approach: we exponentially reduce the number of qubits in the system and then carefully select the appropriate measurement bases to achieve additional savings in both qubit count and measurement resources.
Such logarithmic reduction effectively compresses the original Hilbert space into a smaller, yet information-preserving subspace, ensuring that the essential dynamical and spectral properties of the Hamiltonian remain intact.
When combined with optimized measurement strategies, this approach allows for an even more resource-efficient estimation of expectation values.
Consequently, it becomes possible to simulate exponentially larger systems using only a modest number of physical qubits, substantially extending the reach of variational quantum algorithms on near-term devices.

Using a standard binary encoding, each of the \(N\) SES basis states is mapped
to an \(n\)-qubit register with \(n=\lceil \log_{2} N\rceil\). The site label
\(k\in\{0,\ldots,N-1\}\) is written in binary as a bit string
\(b(k)=(b_1,\ldots,b_n)\), and the corresponding register state is
\(|b_1\cdots b_n\rangle=\big|\mathrm{binary}(k)\big\rangle\).

This encoding strategy is most natural when the number of qubits matches an
exact power of two, i.e., when $N = 2^{n}$. In this setting, each computational
basis state of the $n$-qubit register corresponds directly to one of the $N$
sites of the system, and no additional restructuring is required. When the
system size lies between two consecutive powers of two, $2^{n-1} < N < 2^{n}$,
the situation becomes a bit more complicated: the Hilbert space of the register then
contains more basis states than are actually needed to represent the system.
If left untreated, a variational algorithm would naturally explore this larger
space, which may lead to parameter updates that leave the physical domain.

There are two possible strategies for addressing this mismatch:  
(i) modify the cost function by constructing a new, artificial Hamiltonian energetically punishing the newly introduced states, or  
(ii) restrict the variational circuit (ansatz) so that the evolution is
confined to the subset of states corresponding to the true system's degrees of
freedom.

\subsection{Extended Hamiltonian with generic ansatz}\label{sec:exd_Hamiltonian}

The first approach involves enlarging the Hilbert space and extending the Hamiltonian accordingly.  
This is achieved by adding penalty terms that energetically suppress any population outside the physical subspace:
\begin{equation}
    H_{\mathrm{ex}} = H_o + \sum_{i=N+1}^{2^n} C_p \,\frac{1-Z_i}{2}.
\end{equation}
Here, $H_o$ denotes the original Hamiltonian, and $C_p$ is a large positive constant chosen such that any non-physical configuration incurs a prohibitively high energy cost.  
As a result, during variational optimization, amplitudes in the unphysical subspace are strongly suppressed, while the low-energy spectrum of the physical sector remains effectively unaltered.  
In practice, $C_p$ must be selected significantly larger than the characteristic energy scales of $H_o$, ensuring that the optimizer has no incentive to explore forbidden regions of the Hilbert space.  
Once this extended Hamiltonian is constructed, one can proceed with standard reduction techniques as if the system naturally possessed $2^n$ valid states.

A key advantage of this method is that it allows the use of any $n$-qubit ansatz, such as a hardware-efficient ansatz \cite{kandala2017hardware}, without imposing additional restrictions on the parameter space, since the penalty terms automatically exclude unphysical states from the optimization process.
The price to pay is that a significantly (up almost double) Hilbert space needs to be researched, leading to an increased risk of redundant parameter symmetries, large barren plateaus, and poor gradient scaling, all of which hinder convergence of the optimization process. Even more importantly, all relevant hardware-efficient ansätze are biased in the sense that they naturally favor less complex states (as they start from a product state and use only a very limited number of entangling operations), while a large part of the Hilbert space remains inaccessible \cite{note_HEA_bias_hilbert}. 

As there is no one-to-one physical correspondence between states in the
reduced system and states in the original system (where all states are, in a
sense, comparable), the algorithm may favor ``simple'' solutions over more
complex ones, potentially leading to biased or incorrect results.

\subsection{Binary encoded SES ansatz} \label{sec:reduced SES ansatz}
The second approach avoids modifying the Hamiltonian and instead enforces the restriction at the level of the variational ansatz.  
The idea is to construct a parameterized quantum circuit whose support lies strictly within the physical subspace, such that amplitudes for non-physical states are identically zero throughout the optimization.  
Although this requires more careful circuit design, it has several attractive features.  
In particular, the ansatz can be engineered to preserve the same parameterization as the original ansatz defined for the unreduced Hamiltonian, which means that the numerical optimization landscape--and thus the physical results--remain consistent with those obtained from the full system.

%---------------------------------------------------

We design a quantum circuit that operates within a logarithmically reduced Hilbert space using a combination of data and two ancillary qubits. The key idea is to transfer the amplitude encoded in a pair of ancillary qubits using the $A$ gate (Eq. \ref{fig:Agate}), to the appropriate data-register basis state, thus emulating the original single-excitation ansatz Fig. \ref{fig:ansatz} without ever populating non-physical subspace.

To effectively reproduce the same state as generated by the original ansatz, we employ two ancillary qubits for the application of the $A$ gate.
Conditioned on one of these ancillary qubits, which encodes the amplitude information, we prepare the single-excitation subspace state in the data register.
After this step, the ancillary qubit is reset, and the procedure is repeated: the $A$ gate is applied again to regenerate the amplitude information, which is then transferred to the data register through a sequence of controlled operations. This can be summarized into the following steps:
\begin{enumerate}
\item
\textit{Preparation of $\ket{1}$}: Only once, at the beginning of the procedure, the first qubit is initialized.

\item
\textit{Application of $A$ gate}: The two ancillary qubits are updated via an $A$ gate, which performs a controlled two-qubit rotation that mixes the ancillary states.

\item 
\textit{State preparation on data register}: Controlled on the first ancilla qubit, the appropriate computational basis state $|i\rangle$ is prepared on the (logarithmically reduced) data register using binary encoding.

\item 
\textit{Unflagging ancilla qubit}: A multi-controlled Toffoli gate is used to reset the ancilla back to $|0\rangle$, conditioned on the data register being in state $|i\rangle$.
\end{enumerate}

\begin{figure*}[t]
\centering
\resizebox{\textwidth}{!}{%
\begin{quantikz}[row sep=0.35cm, column sep=0.55cm]
% -------------------- headers --------------------
\lstick{$a_0$} & \gate{X} & \ldots
  % ===== Module 1 (cols 2..5) =====
  & \gate[2]{A_i}
    \gategroup[wires=3,steps=4,style={dashed,rounded corners,fill=blue!20,inner xsep=2pt},
               background,label style={label position=below,anchor=north,yshift=-2pt}]{\sc Module 1}
  & \permute{2,1} & \qw           & \qw
  % ===== Module 2 (cols 6..9) =====
  & \gate[2]{A_{i+1}}
    \gategroup[wires=3,steps=4,style={dashed,rounded corners,fill=blue!20,inner xsep=2pt},
               background,label style={label position=below,anchor=north,yshift=-2pt}]{\sc Module 2}
  & \permute{2,1} & \qw           & \qw  & \dots\\
\lstick{$a_1$} & \qw & \ldots
  % ===== Module 1 =====
  &             & \targX{}   & \ctrl{1} & \targ{}
  % ===== Module 2 =====
  &             & \targX{}   & \ctrl{1} & \targ{} & \ldots\\
\lstick{\text{data}} & \qwbundle{n} & \ldots
  % ===== Module 1 =====
  & \qw         & \qw        & \gate{|i\rangle_{\mathrm{prep}}} & \ctrl{-1}
  % ===== Module 2 =====
  & \qw         & \qw        & \gate{|i\!+\!1\rangle_{\mathrm{prep}}} & \ctrl{-1} & \ldots
\end{quantikz}
}%
\caption{Two concatenated modules of the subspace-constrained ansatz. 
In each module, a two-qubit gate \(A\) acts on the flag–selector pair \((a_0,a_1)\), followed by a SWAP. 
Conditioned on the flag, the data register is prepared in \(|i\rangle\) (Module~1) and \(|i{+}1\rangle\) (Module~2), and a multi-controlled toffoli operation targets \(a_1\).}
\label{fig:ansatz_lr}
\end{figure*}
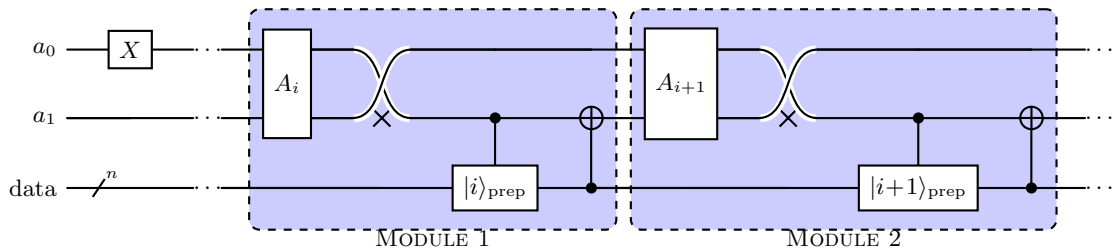

The points 2 to 4 are repeated for each $i<N$, sequentially building up a coherent superposition on the data register. This procedure is illustrated in the Fig. \ref{fig:ansatz_lr}.
This construction guarantees that only valid states from the single-excitation subspace are ever populated. 

To efficiently implement this ansatz we use shifted binary encoding. Concretely, each site label
\(k\in\{0,\ldots,N-1\}\) is stored as the binary string of \(k+1\) (mod
\(2^n\)), i.e.\ \(\lvert k\rangle \mapsto \lvert b_1 b_2\cdots b_n\rangle
=\lvert \mathrm{binary}(k+1)\rangle\).  This choice makes the initialized data
register \(\lvert 0\cdots0\rangle\) distinct from the first prepared state
(\(k=0\) encodes to \(\lvert 0\cdots01\rangle\)).  For instance, an
eight-state SES uses three qubits with the mapping
\(k=0\to\lvert001\rangle,\;k=1\to\lvert010\rangle,\ldots,\;k=7\to\lvert000\rangle\),
as summarized in Table \ref{tab:ses-mapping}.
\begin{table}[ht]
  \centering
  \scriptsize
  \resizebox{0.6\columnwidth}{!}{%
    \begin{tabular}{c c c}
      \toprule
      SES state \(k\) & Basis \(\ket{\,\cdot\,}\) & Binary \(\ket{b_1b_2b_3}\) \\
      \midrule
      0 & \(\ket{00000001}\) & \(\ket{001}\) \\
      1 & \(\ket{00000010}\) & \(\ket{010}\) \\
      2 & \(\ket{00000100}\) & \(\ket{011}\) \\
      3 & \(\ket{00001000}\) & \(\ket{100}\) \\
      4 & \(\ket{00010000}\) & \(\ket{101}\) \\
      5 & \(\ket{00100000}\) & \(\ket{110}\) \\
      6 & \(\ket{01000000}\) & \(\ket{111}\) \\
      7 & \(\ket{10000000}\) & \(\ket{000}\) \\
      \bottomrule
    \end{tabular}%
  }
  \caption{Mapping of SES basis states to their basis and 3-qubit shifted binary encodings.}
  \label{tab:ses-mapping}
\end{table}

%-----------------------------------------------------

The resource need of the proposed ansatz is governed by the number of modules needed to span the
single–excitation subspace (SES) and by the cost of each module. Starting from \(N\) SES basis
states, we encode the excitation index into \(n=\lceil\log_{2}N\rceil\) qubits; one module is invoked
per valid label, for a total of \(N\) modules. %(equal to \(2^{n}\) in the perfectly packed case).

Step~2 applies the $A$ gate on the two ancillary qubits, which can be decomposed into three CNOT gates as shown in Fig. \ref{fig:Agate}. 
%Since this cost is constant and independent of the data-register size, it is ignored in the asymptotic scaling analysis. 
In Step~3, the desired computational basis state $\ket{i}$ is prepared on the data register, controlled on a single ancilla qubit. 
This corresponds to a single-control, multi-target $X$ operation, decomposable on $O(n)$ (and on average $n/4$) CNOT gates. 
%Preparing an $n$-qubit state in this manner requires up to $(n-1)$ CNOTs (and on average $n/2$), giving a scaling of $O(n)$ per module. 
The last step, unflagging the ancilla, is the most resource-intensive. It requires implementing an $n$-controlled Toffoli gate conditioned on the data register. With one additional clean ancilla, such an $n$-controlled $X$ can be realized using $(2n-3)$ Toffoli-type blocks~\cite{khattar2025rise}. 
Each of these blocks can be implemented as a standard $\mathrm{CCX}$ gate, costing six CNOTs~\cite{10.5555/2011791.2011799}. 
Hence, the CNOT cost in this step scales linearly with $n$, i.e., $O(n)$. 
%In contrast, without an additional ancilla, the best-known constructions scale quadratically with $n$, i.e., $O(n^2)$~\cite{barenco1995elementary,shende2008cnot}.
%Consequently, in the worst-case scenario, 
All in all, the total CNOT cost per module scales as $O(n)$. 
As the full ansatz consists of $N$ modules, the overall cost scales as
$O(N n) = O\!\big(N (\log N)\big)$.

\section{Measurement settings}

In our previous work~\cite{krejvci2025minimum}, for a general SES ansatz we 
showed that the measurement protocol can be implemented in an optimal 
manner: all information needed to evaluate the SES Hamiltonian can be
obtained using only three global measurement settings, rather than the 
$\mathcal{O}(N)$. 
Here, we present a subtly different, yet equivalent, formulation that will be convenient for the upcoming subsection \ref{mesurement-binary} on binary-encoded measurement protocol. For clarity, we briefly revisit the correlators involved and the organization of measurements at the level of global settings. This reformulation preserves the three-setting measurement cost and makes explicit the structure we later exploit for implementing efficient measurement protocol in the binary encoded system.

For SES Hamiltonins expressed in Eq. \eqref{H}, the correlators \(\langle X_jY_k\rangle\) and \(\langle Y_jX_k\rangle\) differ only by sign, and thus can also be combined into a single measurement setting. More precisely,
\begin{equation} \label{XX}
    \langle X_jX_k\rangle = \langle Y_jY_k\rangle = 2 |\alpha_j| |\alpha_k| \cos(\theta_k-\theta_j),
\end{equation}
where $|\alpha_j|$, $|\alpha_k|$ and $\theta_j$, $\theta_k$ denote the magnitudes and phases associated with the basis states $\ket{e_j}$ and $\ket{e_k}$.  

Similarly,  
\begin{equation} \label{XY}
    \langle X_jY_k\rangle  = 2 |\alpha_j| |\alpha_k| \sin(\theta_k-\theta_j),
\end{equation}
\begin{equation} \label{YX}
     \langle Y_jX_k\rangle = - 2 |\alpha_j| |\alpha_k| \sin(\theta_k-\theta_j).
\end{equation}

Eqs.~\eqref{XX}--\eqref{YX} show that the Hamiltonian’s expectation value can be fully determined once, for every pair of basis states $\ket{e_j}$ and $\ket{e_k}$, one knows the absolute magnitude of their overlap together with the cosine and sine of their relative phase.

Using the relations in Eqs.~\eqref{XX}--\eqref{YX}, and as described in Appendix~\ref{app:ham-exp}, one can evaluate the cost function or Hamiltonian expectation value as

\begin{align} \label{eq-energy}
E &= \sum_{k=l}^{N} h_{kk} |\alpha_k|^2 \notag \\[6pt]
&\quad + \sum_{j=1}^{N} \sum_{k>j}^{N} \Bigg[
     (h_{jk}+h_{kj})\,|\alpha_j||\alpha_k| \cos(\theta_k-\theta_j) \notag \\[6pt]
&\qquad\qquad\;
   + \frac{h_{jk}-h_{kj}}{2i}\,|\alpha_j||\alpha_k| \sin(\theta_k-\theta_j)
   \Bigg].
\end{align}

\subsection{Measurement protocol} \label{meas_pro}
% Taking these observations into account--together with the single-qubit \(Z_j\) terms--one finds that the full expectation value of the Hamiltonian can be computed if the amplitudes \(|\alpha_j|\) and relative phase differences \(\theta_k - \theta_j\) between each pair of sites are known. 
The extraction of amplitudes and phases from quantum measurements follows naturally from the structure of the variational state in the single-particle picture as decribed below.

\subsubsection{Obtaining amplitudes from $Z$-measurements}

In particular, determining the site occupations--or equivalently, the magnitudes of the expansion coefficients--is a straightforward first step, as it relies solely on measurements in the computational basis.
These measurements directly reveal how the probability density of the particle is distributed across the available sites.

In the single-excitation subspace expressed in Eq. \eqref{eq:ses_superposition},
% \begin{equation}
%    |\Psi\rangle=\sum_{j=1}^{N}\alpha_j|e_j\rangle,\qquad 
% \alpha_j=|\alpha_j|e^{i\theta_j}, 
% \end{equation}
the occupation probability of site $j$ is $p_j=|\alpha_j|^2$.  
With computational-basis readout, identifying $|e_j\rangle$ with $Z_j=-1$ and all other qubits with $Z=+1$, the single-qubit expectation satisfies
\begin{equation} \label{amps}
|\alpha_j|=\sqrt{\frac{1-\langle Z_j\rangle}{2}}.
\end{equation}
Knowing all $|\alpha_j|$ fixes the magnitudes.

\subsubsection{Nearest-neighbour phase differences}

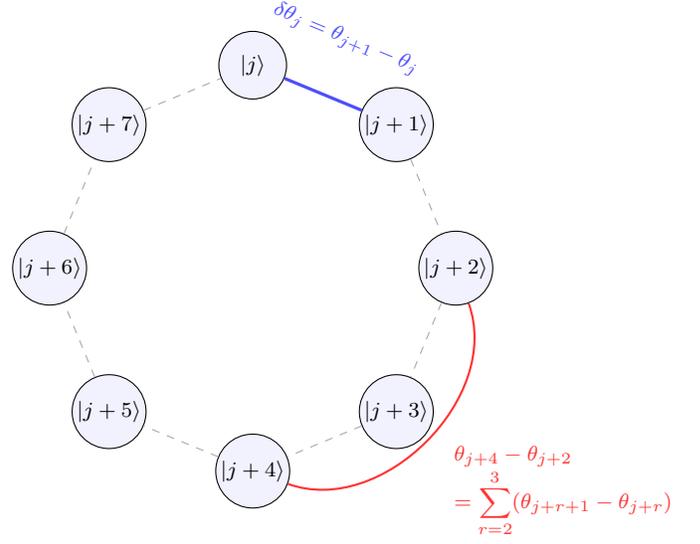
\begin{figure}[h]
\centering
\begin{tikzpicture}[
    scale=0.9,
    state/.style={circle, draw=black, fill=blue!5,
                  minimum size=9mm, inner sep=1pt, font=\footnotesize},
    every node/.style={font=\footnotesize}
]

% Radius of ring
\def\R{3.0}

% --- nodes |j>, |j+1>, ..., |j+7> ---
\node[state]               (s0) at (90:\R)     
  {$\ket{j}$};
\node[state]               (s1) at (45:\R)        {$\ket{j+1}$};
\node[state]               (s2) at (0:\R)         {$\ket{j+2}$};
\node[state]               (s3) at (-45:\R)       {$\ket{j+3}$};
\node[state]               (s4) at (-90:\R)       {$\ket{j+4}$};
\node[state]               (s5) at (-135:\R)      {$\ket{j+5}$};
\node[state]               (s6) at (180:\R)       {$\ket{j+6}$};
\node[state]               (s7) at (135:\R)       {$\ket{j+7}$};

% --- light grey dashed cycle ---
\foreach \a/\b in {s0/s1, s1/s2, s2/s3, s3/s4, s4/s5, s5/s6, s6/s7, s7/s0}
    \draw[dashed, draw=gray!70] (\a) -- (\b);

% --- highlight local delta-theta on j → j+1 with label on the edge ---
\draw[very thick, blue!70]
    (s0) -- node[above, sloped, blue!70, yshift=15pt]
        {$\delta\theta_{j}=\theta_{j+1}-\theta_{j}$}
    (s1);

% --- red arc from j+2 to j+4 with label along the arc ---
\draw[red!80, thick]
    (s2) to[bend right=-65]
      node[midway, below right, red!80, xshift=7pt, yshift=2pt, align=left]
      {$\theta_{j+4}-\theta_{j+2}$\\
       $=\displaystyle\sum_{r=2}^{3}(\theta_{j+r+1}-\theta_{j+r})$}
    (s4);

\end{tikzpicture}
\caption{Circular arrangement of the basis states 
$\ket{j},\ket{j+1},\ldots,\ket{j+7}$ used to extract phase differences. 
The blue edge shows the local relative phase 
$\delta\theta_j=\theta_{j+1}-\theta_j$, while the red arc illustrates the 
cumulative phase difference across multiple steps,
$\theta_{j+4}-\theta_{j+2}=\sum_{r=2}^{3}(\theta_{j+r+1}-\theta_{j+r})$.}
\label{fig:phasecircle}
\end{figure}

Once the amplitudes are known, the next step is to extract phase differences between neighbouring qubits.  
To this end, the qubits are arranged in a cyclic structure  as shown in the Fig. \ref{fig:phasecircle} and nearest-neighbour two-qubit correlators are measured:

\begin{align}
    \langle X_j X_{j+1} \rangle &= 2 |\alpha_j| |\alpha_{j+1}| \cos(\theta_{j+1} - \theta_j), \\
    \langle X_j Y_{j+1} \rangle &= 2 |\alpha_j| |\alpha_{j+1}| \sin(\theta_{j+1}- \theta_j),
\end{align}
where $(j,j+1)$ are neighbouring sites.  
The relative phase difference between basis states $j$ and $j+1$ is then directly obtained as
\begin{equation}
    \theta_{j+1} - \theta_j = \arctan\!\left( \frac{\langle X_j Y_{j+1} \rangle}{\langle X_j X_{j+1} \rangle} \right).
    \label{eq:phase_diff}
\end{equation}

%-------------------------------------------

%--------------------------------------

\subsubsection{Phase differences between non-neighbouring qubits}

For qubits that are not directly connected, relative phase differences can be reconstructed by summing over paths through the cyclic graph as shown in the Fig. \ref{fig:phasecircle}.  
For example, if $j$ and $k$ are separated by intermediate qubits $j+1, j+2, \dots, k-1$, then
\begin{equation}\label{eq:off_neigh}
    \theta_k - \theta_j = \sum_{m=j}^{k-1} \bigl(\theta_{m+1} - \theta_m \bigr),
\end{equation}
where each nearest-neighbour difference $\theta_{m+1} - \theta_m$ is obtained from Eq.~\eqref{eq:phase_diff}.  

\paragraph*{Global measurement settings}:

In summary, the full reconstruction requires only three global measurement settings to obtain 
the amplitudes $|\alpha_i|$ and all relative phases $\theta_j - \theta_k$ for arbitrary $j,k$:  

\begin{itemize}
    \item $\mathcal{M}_Z$: One round of measurements in the computational basis, yielding
    \[
        |\alpha_j|, \; |\alpha_j|^2, \quad \text{for all } j=1,\dots,N.
    \]

    \item $\mathcal{M}_{XX}$: One round of measurements in the $X$ basis, yielding nearest-neighbour correlations
    \[
        \langle X_j X_{j+1} \rangle, \quad \text{for neighbouring pairs } (j,j+1).
    \]

    \item $\mathcal{M}_{XY}$: One round of measurements in an alternating pattern of $X$ and $Y$, e.g.\ $(X,Y,X,Y,\dots)$, yielding nearest-neighbour mixed correlators
    \[
        \langle X_j Y_{j+1} \rangle, \quad \text{for neighbouring pairs } (j,j+1).
    \]
\end{itemize}

From these three settings, all nearest-neighbour phase differences are obtained via Eq.~\eqref{eq:phase_diff},  
and phase differences between arbitrary qubit pairs follow by summing along paths in the cycle.  
Thus, amplitudes and relative phases of the entire state in the single-excitation subspace can be fully reconstructed using only three measurement settings.

\subsubsection{Handling vanishing amplitudes}

The difficulty with this protocol arises when Eq.~\eqref{eq:phase_diff} fails, i.e., when one of the amplitudes $|\alpha_j|$ or $|\alpha_k|$ vanishes for a given pair $(j,k)$.
In such cases, both the numerator and denominator of the ratio become zero, leaving the angle undefined and thereby obstructing the calculation of phase differences along paths involving that pair.
To address this, one can simply omit states with vanishing amplitudes ($|\alpha_j|\approx 0$) and reconstruct the cyclic graph using only non-vanishing states.
This is reasonable, as states with negligible amplitudes contribute insignificantly to the expectation value of the Hamiltonian.

Any qubit with close to zero amplitude simply forms a disconnected arm of the circle and can be ignored, since such qubits contribute nothing to the total energy and do not interfere with the determination of relative phases among the remaining qubits.

The measurement strategy introduced here differs from the one
in our earlier work~\cite{krejvci2025minimum}, where the phase differences were
extracted in an alternative, but equivalent, way.  There, the relative phase
between basis states was encoded in the complex two–qubit correlator
\begin{equation}
  C_{jk}
  = \langle X_j X_k \rangle
    + i \langle X_j Y_k \rangle
  = 2\,|a_j|\,|a_k|\,e^{i\varphi_{jk}},
  \tag{15}
\end{equation}
whose argument directly yields the phase difference $\varphi_{jk}$ between sites
$j$ and $k$.  
Using this quantity, phase differences for opposite–parity pairs
$(j,k)$ can be determined from just the $\mathcal{M}_{XX}$ and
$\mathcal{M}_{XY}$ global settings.  
Once those odd–parity phases are
fixed, the remaining even–parity relations follow recursively from the odd-parity relations, so that all
relative phases are recovered without introducing any additional measurement
settings.  For more detailed discussion on recursive construction of phase differences, please refer to  the article~\cite{krejvci2025minimum}.

%--------------------------------------------------------

\begin{figure}[t]
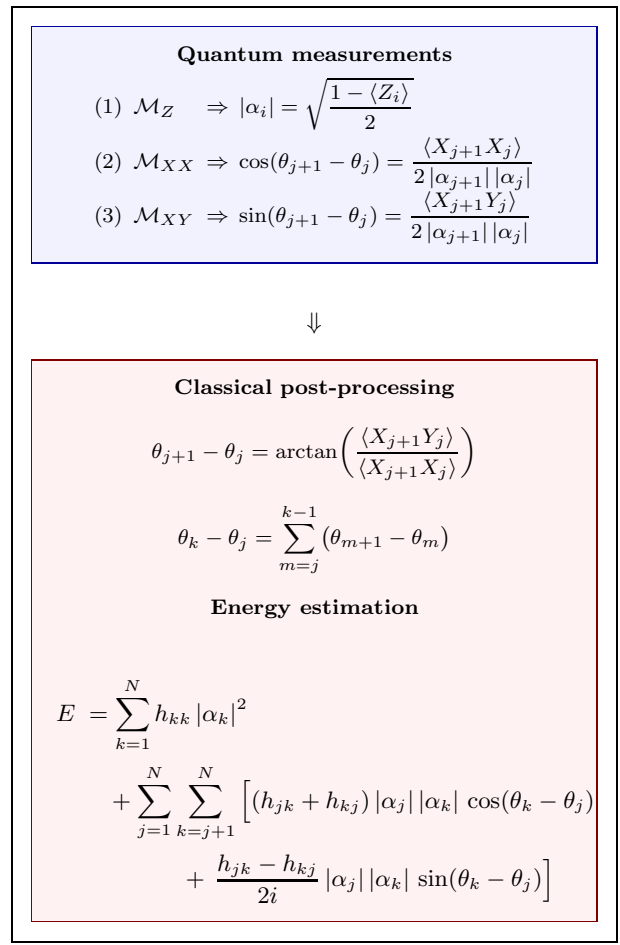

\centering
\begin{tcolorbox}[
    width=0.8\textwidth,
    colback=white, colframe=black,
    boxrule=0.8pt, sharp corners,
    left=4pt, right=4pt, top=4pt, bottom=4pt
]

% =======================
% Quantum Measurement Box
% =======================
\begin{tcolorbox}[
    colback=blue!5!white, colframe=blue!60!black,
    boxrule=0.6pt, sharp corners,
    left=6pt, right=6pt, top=4pt, bottom=4pt
]
\footnotesize
\textbf{Quantum measurements} \\[4pt]
\begin{tabular}{@{}rlcl@{}}
(1) & $\mathcal{M}_{Z}$  & $\Rightarrow$ & $\displaystyle |\alpha_i| = \sqrt{\frac{1-\langle Z_i\rangle}{2}}$ \\[6pt]
(2) & $\mathcal{M}_{XX}$ & $\Rightarrow$ & $\displaystyle \cos (\theta_{j+1}-\theta_{j}) = \frac{\langle X_{j+1} X_j \rangle}{2\,|\alpha_{j+1}|\,|\alpha_j|}$ \\[6pt]
(3) & $\mathcal{M}_{XY}$ & $\Rightarrow$ & $\displaystyle \sin (\theta_{j+1}-\theta_{j}) = \frac{\langle X_{j+1} Y_j \rangle}{2\,|\alpha_{j+1}|\,|\alpha_j|}$
\end{tabular}
\end{tcolorbox}

\vspace{2pt}
\begin{center} $\Downarrow$ \end{center}

% =======================
% Classical Processing Box
% =======================
\begin{tcolorbox}[
    colback=red!5!white, colframe=red!50!black,
    boxrule=0.6pt, sharp corners,
    left=6pt, right=6pt, top=4pt, bottom=4pt
]
\footnotesize
\textbf{Classical post-processing} \\[6pt]
\begin{center}
$\displaystyle
\theta_{j+1}-\theta_{j} = \arctan\!\left( \frac{\langle X_{j+1} Y_j \rangle}{\langle X_{j+1} X_j \rangle} \right)
$ \\[8pt]
$\displaystyle
\theta_k - \theta_j = \sum_{m=j}^{k-1} \bigl(\theta_{m+1} - \theta_m \bigr) 
$ \\[8pt]
\textbf{Energy estimation} \\[4pt]
\begingroup
\small
\setlength{\jot}{2pt}
\[
\begin{aligned}
E \; &= \sum_{k=1}^{N} h_{kk}\,|\alpha_k|^2 \\[2pt]
&\quad + \sum_{j=1}^{N} \sum_{k=j+1}^{N}
\Big[
(h_{jk}+h_{kj})\,|\alpha_j|\,|\alpha_k|\,\cos(\theta_k-\theta_j) \\[2pt]
&\qquad\qquad
+\, \frac{h_{jk}-h_{kj}}{2i}\,|\alpha_j|\,|\alpha_k|\,\sin(\theta_k-\theta_j)
\Big]
\end{aligned}
\]
\endgroup
\end{center}
\end{tcolorbox}

\end{tcolorbox}

\caption{The protocol consists of quantum measurements to extract amplitudes and relative phases, followed by classical post-processing to reconstruct the total energy.}
\label{flowchart}
\end{figure}

\subsection{Efficient Measurement Settings in Binary encoded system}\label{mesurement-binary}

In the binary encoding, the first step in characterizing the variational state is to recover the 
amplitudes, $|\alpha_i|$. This task can be accomplished
very simply: a single global measurement of all qubits in the computational ($Z$) basis produces a 
distribution over bitstrings. Since each codeword $\ket{b(j)}$ is uniquely associated with a basis state 
of the single--excitation subspace, the corresponding outcome probability is exactly
\(|\alpha_j|^2\). In other words, the magnitudes of the amplitudes can be directly obtained from the 
frequency statistics of a single all-$Z$ measurement setting. 

The more challenging step is to determine the relative phases between the amplitudes. To achieve this,
we introduce a family of operators that selectively couple two chosen codewords. For any two indices 
\(j,k\), define
\begin{equation}
O_X^{(j,k)} \;=\; \bigotimes_{\ell \in D_{jk}} X_\ell 
\;\;\otimes \!\! \bigotimes_{\ell \in S_{jk}} P^{(b_\ell(j))}_\ell ,
\end{equation}
where the index sets are given by
\begin{align}
D_{jk} &= \{\ell \;:\; b_\ell(j) \neq b_\ell(k)\}, \\
S_{jk} &= \{\ell \;:\; b_\ell(j) = b_\ell(k)\}.
\end{align}
Here $D_{jk}$ denotes the qubits on which the two binary strings $b(j)$ and $b(k)$ differ, while
$S_{jk}$ denotes the qubits where they coincide. On the differing positions, $X$ operators are applied,
whereas on the coinciding positions we project onto the appropriate computational basis state by using
\begin{align}
P^{(0)}_\ell = \frac{1+Z_\ell}{2}, \qquad 
P^{(1)}_\ell = \frac{1-Z_\ell}{2}.
\end{align}
By construction, the operator $O_X^{(j,k)}$ swaps the two codewords $\ket{b(j)}$ and $\ket{b(k)}$ 
while annihilating all other codewords. 

However, this operator is not unique, since different pairs of codewords can correspond to the same pair of difference and sum patterns \((D, S)\).  
In such cases, the operator \(O_X\) couples multiple pairs of basis states simultaneously, and its measured expectation value becomes a sum over several independent contributions rather than isolating a single transition.  
For instance, in a two-qubit encoding, the operator \(O_X = X \otimes X\) acts nontrivially on both the pairs \((00,11)\) and \((01,10)\).  
Consequently, its expectation value contains overlapping information from both transitions, preventing the unique reconstruction of individual relative phases from the measurement outcomes alone.

To eliminate this ambiguity, the operator can be restricted to act only between codeword pairs \((i, j)\) that satisfy \(|D_{ij}| = 1\)--that is, pairs differing by a single bit flip.  
Under this condition, the operator \(O_X\) becomes unique, as each such pair corresponds to a distinct single-qubit transition.  
The measured expectation value then takes the form
\begin{equation} \label{Ox}
\langle O_X^{(j,k)} \rangle \;=\; 2|\alpha_j||\alpha_k|\cos(\theta_k - \theta_j),
\end{equation}
which directly yields the cosine of the relative phase difference between amplitudes \(\alpha_j\) and \(\alpha_k\).  

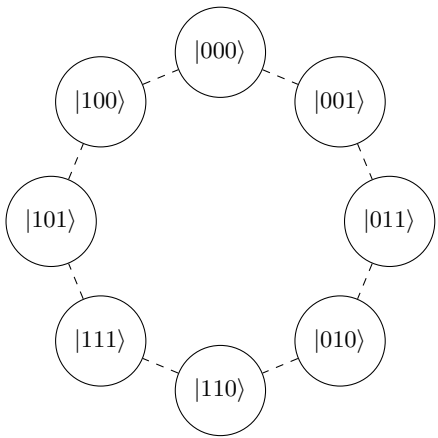
\begin{figure}[h] 
\centering
\begin{tikzpicture}[scale=1.5,
    state/.style={circle, draw, minimum size=12mm, inner sep=0pt, font=\small}]
  % positions on a circle (clockwise from top)
  \path
    (90:1.5)  node[state] (s000) {$|000\rangle$}
    (45:1.5)  node[state] (s001) {$|001\rangle$}
    (0:1.5)   node[state] (s011) {$|011\rangle$}
    (-45:1.5) node[state] (s010) {$|010\rangle$}
    (-90:1.5) node[state] (s110) {$|110\rangle$}
    (-135:1.5)node[state] (s111) {$|111\rangle$}
    (180:1.5) node[state] (s101) {$|101\rangle$}
    (135:1.5) node[state] (s100) {$|100\rangle$};

  % dashed cycle edges
  \draw[dashed] (s000) -- (s001);
  \draw[dashed] (s001) -- (s011);
  \draw[dashed] (s011) -- (s010);
  \draw[dashed] (s010) -- (s110);
  \draw[dashed] (s110) -- (s111);
  \draw[dashed] (s111) -- (s101);
  \draw[dashed] (s101) -- (s100);
  \draw[dashed] (s100) -- (s000);

  % (optional) highlight a segment in red
  %\draw[very thick, dashed, red] (s000) -- (s001);
  %\draw[very thick, dashed, red] (s001) -- (s011);
\end{tikzpicture}
\caption{Cyclic representation of the 3-qubit Gray code sequence, where adjacent states differ by a single bit flip.}
\label{fig:gray_code}
\end{figure}

This uniqueness can be systematically ensured by arranging the codewords along a cyclic \textit{Gray code} path~\cite{gilbert1958gray,Gray1953,Savage1997}, where successive \(n\)-bit strings differ by exactly one qubit flip.  
Following such an ordering establishes a one-to-one correspondence between each operator \(O_X\) and an edge in the Gray code graph, guaranteeing that every measurement isolates a single transition in the encoded space.  
An example of this Gray-coded construction is depicted in Fig.~\ref{fig:gray_code}.

%----------------------------------------
%----------------------------------------

An analogous construction yields the sine terms. If in the definition of the swap operator we
replace $X$ by $Y$ on the differing qubit, we obtain
\begin{equation}
O_Y^{(j,k)} \;=\; \bigotimes_{\ell \in D_{jk}} Y_\ell 
\;\;\otimes \!\! \bigotimes_{\ell \in S_{jk}} P^{(b_\ell(j))}_\ell ,
\end{equation}
whose expectation value gives
\begin{equation} \label{Oy}
   \langle O_Y^{(j,k)} \rangle \; =2|\alpha_j||\alpha_k|\sin(\theta_k-\theta_j).
\end{equation}

Thus, measuring $O_Y$ provides the missing sine information. 

Similar to Eq. \eqref{eq:phase_diff}, from the Eqs. \eqref{Ox}--\eqref{Oy}, one can obtain the phase difference between the neighbouring qubits using the following equation.
\begin{equation}
    \theta_{j+1} -\theta_j = \arctan\!\left (\frac{\langle X_j Y_{j+1}\rangle}{\langle X_j X_{j+1}\rangle}\right).
\end{equation}
The phase difference of the remaining pairs can be calculated by summing along the path of gray code curve, by using the Eq. \eqref{eq:off_neigh}. After which one can calculate the energy of the quantum state for Hamiltonian $H$ using Eq. \eqref{eq-energy}. 

Since there are $n$ possible positions for the
single flip in the Gray code, we need $n$ distinct $X$-based settings for the cosine terms and $n$ 
corresponding $Y$-based settings for the sine terms. In summary, the complete characterization of the encoded variational state requires only $2n + 1$ 
%measurements
%\textcolor{blue}{MF: I suggest to modify the following}
%\begin{equation}
%M_{\mathrm{measurements}} = 2n + 1
%\end{equation}
distinct measurement settings: one global all-$Z$ measurement to determine the magnitudes,
$n$ settings of $O_X$ operators to extract all cosines, and $n$ settings of $O_Y$ operators 
to extract all sines.
This Gray-code--based construction ensures uniqueness of each operator, 
prevents collisions of pairs with same $D$ and $S$, and achieves an efficient scaling of the measurement overhead with the number 
of qubits $n$ in the binary encoded system.
The overall workflow is identical to Fig.~\ref{flowchart}; the only
difference is the cost: $2n+1$ settings instead of three for the SES ansatz.

\section{Resource efficiency}
In classical computation, based on the current hardware architecture consisting of a combination of (large) memory and localized processing unit, the costs of the computation are defined independently by two parameters. One is the size of the data register, defining the minimal size of the computer needed. This limit is unavoidable and, for example, also limiting the size of a quantum state that can be simulated classically. The other parameter is the number of operations that need to be performed, defining the time costs of the algorithm. For deterministic programs one run suffices, thus even a computer with large data register cannot help to reduce the time costs of an algorithm with high number of operations needed. 

For quantum computers the situation is significantly different, as summarized below:
\begin{itemize}
    \item While the size of a quantum computer is a similar limitation than in classical space, one can in principle encode exponentially more information into the same size of the device
    \item As in most existing scalable quantum infrastructure the number of possible parallel operations scales with the number of qubits, the depth (number of layers) of the quantum circuit is more relevant than the actual number of operations to be performed.
    \item As quantum computers are stochastic by nature, runs must be repeated and
    statistics gathered from measurements. Consequently, increasing the size of the
    device can reduce runtime via
    parallelization.
 
    %As quantum computers are stochastic by its nature, there running needs to be repeated and statistics read. Thus the size of the computer can serve to decrease the running time by parallelization
    
    \item Unlike the classical case, a single measurement will not reveal all the information about the system. Thus the number of measurement settings needed inherently enters into the time costs of the algorithm, as it increases the running time.
  
\end{itemize}

Based on these differences, we introduce a volumetric efficiency metric for quantum devices
\begin{align}
    \mathcal{E} \;=\; 
    (\text{qubit width}) \times (\text{circuit depth}) \times \\ \nonumber
    (\text{number of measurement settings}),
\end{align}
which captures the total space–time–sampling volume that a hardware platform
must support during its run. This measure is especially well suited for VQE applications, as all parameters of the circuit (size, depth and measurement settings) can be adjusted. 

Let us now summarize the results for the above presented algorithms. 
The original single-excitation encoding uses $N$ qubits, depth
$D_{\mathrm{SES}} = O(N)$ from the linear chain of two-qubit gates by application of A gates, and
three global measurement settings for extracting amplitudes and relative
phases.  The corresponding volumetric cost thus scales as
\begin{equation}
    \mathcal{E}_{\mathrm{Original}}
    \;=\;
    N \times O(N) \times 3
    \;=\;
    O(N^{2}).
\end{equation}

In the logarithmic-qubit encoding presented above, the width reduces to 
%$n = \lceil \log_{2} N \rceil$ 
$n$ qubits.  
The depth is different depending on the approach chosen. For hardware efficient ansatz, it can be as low as $O(n)$, while for the most general preparation scheme~\cite{plesch2011quantum} it is $O(N)$ and when using the preparation mimicking original SES parametrization it increases to 
 $O(N n)$.
The number of global measurement settings is then independent on the preparation protocol and is
$2n+1 = O(n)$. 
Thus the overall volumetric cost becomes
\begin{align}
    \mathcal{E}_{\mathrm{HE}}
    \;=\;
    n \times O(n) \times O(n)=O\!\left( (\log N)^{3} \right) \\ \nonumber
   \mathcal{E}_{\mathrm{Full}} \;=\;
    n \times O(N) \times O(n)=O\!\left( N (\log N)^{2} \right) \\ \nonumber
    \mathcal{E}_{\mathrm{Gray}} \;=\;
    n \times O(N n) \times O(n)=O\!\left( N (\log N)^{3} \right) .
\end{align}
These results are depicted also in Tab. \ref{tab:resource-scaling}. 

Consider two illustrative qubit sizes of interest. For \(N=1024\) sites, plausibly reachable on
near-term hardware, the binary-encoded approach yields approximate speedups relative to the original SES implementation of \(1000\) with a hardware-efficient (HE) ansatz and \(10\) with a general ansatz over the full Hilbert space, while the
ansatz that mimics the original SES preparation has essentially the same
volumetric cost as the original SES.

On the other hand, for $N = 1 048 576$, already of  practical interest, the corresponding approximate speedups
relative to the original SES implementation are
\( 10^{8}\) for a hardware–efficient ansatz,
\( 2.6\times 10^{3}\) for a general ansatz over the full Hilbert space,
and \( 130\) for the SES–mimicking ansatz.
For intuition, a computation that would take about a year with the
original SES mapping could be reduced to a few days with the SES–mimicking
ansatz, a few hours with a general ansatz, and to a fraction of a second with a
hardware-efficient ansatz--although here the chance of a failure would probably be rather high. These estimates also ignore hardware size: the original mapping
would require on the order of a million physical qubits, whereas our binary
encoding needs only \(n=\lceil\log_{2}N\rceil=20\) qubits in this example. 

\begin{table*}[t]
\small
\centering
\begin{tabular}{lcccc}
\toprule
\textbf{Resource} & \textbf{Original SES} &
\multicolumn{3}{c}{\textbf{Binary-encoded SES}} \\
& & \textit{Hardware-efficient} & \textit{Full} & \textit{Gray-code} \\
\midrule
Qubits                       & $N$ & \multicolumn{3}{c}{$n=\lceil \log_2 N\rceil$} \\
Measurement settings         & $3$ & \multicolumn{3}{c}{2$n$+1}\\
Circuit depth (CNOT)         & $\mathcal{O}(N)$ & $\mathcal{O}(n)$ & $\mathcal{O}(N)$ & $\mathcal{O}(Nn)$ \\
\addlinespace[2pt]
\textbf{Volumetric cost} $\;\mathcal{E}$ 
& $\mathcal{O}(N^2)$
& $\mathcal{O}\!\big((\log N)^{3}\big)$
& $\mathcal{O}\!\big(N(\log N)^{2}\big)$
& $\mathcal{O}\!\big(N(\log N)^{3}\big)$ \\
\bottomrule
\end{tabular}
\caption{Comparison of asymptotic resource scaling. The three Binary-encoded SES
columns correspond to (i) Hardware-efficient ansatz,
(ii) a general preparation scheme, and (iii) our Gray-code
measurement scheme. Here $N$ is the number of SES basis states and
$n=\lceil \log_2 N\rceil$ is the number of qubits in the compressed register.}
\label{tab:resource-scaling}
\end{table*}

\section{Conclusion}

We have introduced an efficient framework for implementing solid-state Hamiltonians on quantum hardware by encoding the physical subspace into an exponentially smaller number of qubits.  
By reformulating the single-particle problem within a binary-encoded register and designing an ansatz that preserves the parameterization of the original SES circuit, we achieved an accurate representation of the physical wavefunction using only about $\log_2 N$ qubits.  
The proposed encoding compresses the Hilbert space exponentially while remaining fully compatible with variational algorithms such as the VQE, thus enabling simulations of larger systems on near-term quantum hardware.

From a resource standpoint, the logarithmic-qubit encoding adds only a
polylogarithmic overhead in circuit depth and measurement settings, while
reducing the qubit width exponentially. The resulting volumetric cost
%$\mathcal{E} = (\text{width}) \times (\text{depth}) \times (\text{settings})$,
improves significantly,
%$\eta_{\mathrm{vol}} \sim N/(\log N)^3$
meaning that large 
Hamiltonians can be simulated on dramatically smaller registers with comparable
circuit complexity.

The proposed scheme provides a practical, hardware-efficient route
for simulating solid-state Hamiltonians in logarithmically reduced qubit
spaces. The same principles--compact encodings, structured ansätze, and
measurement-efficient protocols--can be extended to more complex systems such as two-excitation Hamiltonians, as long as they occupy only a small fraction of the whole Hilbert space, allowing scalable variational simulations on near-term quantum devices. Studying noise robustness, extensions to correlated multi-electron sectors, as well as the resulting measurement complexity in compact encodings are promising directions for future work.

\begin{acknowledgments}
The authors acknowledge Ivana Miháliková's helpful discussions and feedback.

M.F. acknowledges financial supports from (i) the Czech Academy of Science (the {\it Praemium Academiae} and the Strategy AV21, in particular the program ``AI: Artificial Intelligence for Science and Society") and (ii) the Ministry of Education, Youth and Sports of the Czech republic (project No. LUC25028  within the INTER-EXCELLENCE II program, a subprogram INTER-COST - LUC25).
I.A.M. acknowledges the support of the VEGA project No. 2/0055/23 and Research and Innovation Authority project 09I03-03-V04-00685. M.P. acknowledges the support of project Research and Innovation Authority project 09I03-03-V04-00425.
\end{acknowledgments}

\onecolumn
\appendix

\section{Expectation of the Hamiltonian in the SES} \label{app:ham-exp}

Consider an SES state of the form
\begin{equation}
|\Psi\rangle = \sum_{k=1}^{N} \alpha_k |k\rangle, 
\qquad \alpha_k = |\alpha_k| e^{i\theta_k},
\end{equation}
where $|k\rangle$ denotes a single-excitation basis state and 
$\alpha_k$ is expressed in polar form with modulus $r_k = |\alpha_k|$ 
and phase $\theta_k$.

The Hamiltonian restricted to this subspace is
\begin{equation}
H = \sum_{k} h_{kk} |k\rangle \langle k|
  + \sum_{j \neq k} h_{jk} |j\rangle \langle k|, 
\qquad h_{kj} = h_{jk}^* .
\end{equation}

The expectation value of $H$ is
\begin{eqnarray}
E &=& \langle \Psi| H | \Psi \rangle \nonumber \\
  &=& \sum_{k} h_{kk} |\alpha_k|^2 
   + \sum_{j \neq k} h_{jk}\,\alpha_j^* \alpha_k. \qquad (A1)
\end{eqnarray}

Since each off-diagonal term appears twice, we group them as
\begin{equation}
E = \sum_{k} h_{kk} |\alpha_k|^2
  + \sum_{j<k} \left( h_{jk}\,\alpha_j^*\alpha_k 
  + h_{kj}\,\alpha_k^*\alpha_j \right).
\end{equation}

Now substitute $\alpha_j^*\alpha_k = r_j r_k e^{i(\theta_k-\theta_j)}$ 
and define the relative phase difference $\Delta_{jk} = \theta_k - \theta_j $. This gives
\begin{eqnarray}
h_{jk} e^{i\Delta_{jk}} + h_{kj} e^{-i\Delta_{jk}}
&=& (h_{jk}+h_{kj})\cos \Delta_{jk} \nonumber \\
&& - \frac{h_{jk}-h_{kj}}{i}\,\sin \Delta_{jk}. 
\end{eqnarray}

Using Hermiticity ($h_{kj}=h_{jk}^*$), this separates into real and imaginary parts.  
The final compact expression for the expectation value is
\begin{eqnarray}
E &=& \sum_{k=1}^{N} h_{kk} |\alpha_k|^2 \nonumber \\
&& + 2 \sum_{j=1}^{N} \sum_{k>j}^{N}
   \Big[ \mathrm{Re}(h_{jk})\,|\alpha_j||\alpha_k|\cos(\theta_k-\theta_j) \nonumber \\
&& \qquad - \mathrm{Im}(h_{jk})\,|\alpha_j||\alpha_k|\sin(\theta_k-\theta_j) \Big]. 
\end{eqnarray}

This is the desired expression, with diagonal terms from $h_{kk}$ 
and off-diagonal contributions separated into real and imaginary parts of $h_{jk}$.

\bibliographystyle{unsrtnat}
\bibliography{bibliography_doi}

\begin{thebibliography}{36}
\bibitem[not()]{note_HEA_bias_hilbert}
A generic pure state on $n$ qubits lives in a an exponentially big Hilbert space and, up to normalization and global phase, requires $\mathcal{O}(2^n)$ real parameters to specify. By contrast, a hardware-efficient ansatz (HEA) with fixed layout and limited depth contains only $\mathcal{O}(p n)$ continuous parameters (for $p$ layers) and uses a restricted set of entangling operations. Starting from a simple product state such as $|0\rangle^{\otimes n}$ and applying a shallow, architecture-constrained circuit therefore explores only a tiny subset of the full state space. This induces an intrinsic bias toward low-complexity states and leaves most of Hilbert space effectively inaccessible at practical depths.
\end{thebibliography}

\end{document}